\documentclass[aps,prl,twocolumn,groupedaddress]{revtex4-1}

\usepackage{graphicx}

\begin{document}


\title{Carpet cloaking and Laplace transformation}


\author{T. Ochiai}

\email[]{ochiai@otsuma.ac.jp}
\affiliation{School of Social Information Studies, Otsuma Women's University, 2-7-1 Karakida, Tama-shi, Tokyo 206-8540, Japan}

\author{J.C. Nacher}

\email[]{nacher@fun.ac.jp}
\affiliation{
Department of Complex and Intelligent Systems, Future University Hakodate,  116-2 Kamedanakano-cho, Hakodate, 041-8655, Hokkaido, Japan}


\date{\today}

\begin{abstract}
Recently, researchers have proposed several carpet 
cloaking designs that are able to hide a real object under a bump in a way that it is perceived as a flat ground plane. 
Here, we present a method to design two-dimensional isotropic carpet 
cloaking devices using Laplace transformation. We show that each functional form of a Laplace transformation corresponds 
to a different carpet cloaking design. Therefore, our approach allows us to 
systematically design a rich variety of cloaking devices.
Our analysis includes several examples containing different bump geometries 
that illustrate the proposed methodology.
\end{abstract}


\maketitle

\section{Introduction}

For many years, the design and realization of cloaking devices (i.e. invisibility devices) have been one of the scientists dreams 
and have inspired a number of movies and novels. Cloaking devices are intended to guide light rays 
to go around an object. As a result, the object will be hidden from sight \cite{exp3}. 
 
Recently, advances in transformation optics and metamaterials science have opened new avenues to 
control electromagnetic waves at will and have enabled reseachers to propose for the first time several designs of cloaking devices 
\cite{exp3, pendry2006, exp1, exp2, leon1, leon2, jiang2, our, our2}. 
Though theoretically such devices 
can be designed, mathematical analyses have shown that extreme values of permittivity $\epsilon$ and permeability $\mu$
are necessary to realize the cloaking structure. Metamaterials, however,
are offering a promising solution for this issue, though with still limited succcess. 
 
Metamaterial is a man-made material that consists of periodic geometric and metallic structures. When an electromagnetic wave 
enters a metamaterial, it behaves as a $LC$ circuit that induces a response to the incident electromagnetic waves \cite{meta}.
The development of these materials have led to new possible applications in several areas such as 
medical imaging, nanotechnologies, telecommunications and defense. When the size of each periodic structure is smaller than 
the incoming wavelength, the metamaterial behaves as a continuous optical material with an effective 
permittivity and permeability. A unique optical property of metamaterials is that not only a relatively high (or low) $\epsilon$ and $\mu$ can be 
obtained, but also that negative refraction phenomena is also possible \cite{meta, our, our2}, which can not be found in nature. Though 
metamaterials could be applied 
to a new kind of systems like perfect 
lenses \cite{pendry2000} and cloaking devices \cite{report}, the neccesity of extreme values of $\epsilon$ and $\mu$
makes difficult the realization of these devices.
 
Researchers have recently proposed a new type of cloaking device called {\it carpet cloaking} that avoids the problem 
of extreme values of $\epsilon$ and $\mu$ \cite{li, liu, ergin}.  Carpet 
cloaking is a flat mirror-like device which has a bump at the center. However, an observer will not see the bump and, instead of that, 
will only perceive a flat ground plane. This effect is achieved because the metamaterial can properly guide the light rays around a bump.
A remarkable feature of the carpet cloaking is that it only requires a reasonable range of positive refractive index 
profile that can be achieved with current materials science technology. 

Several types of carpet cloakings have been recently proposed \cite{zhang} and some experiments 
for carpet cloaking have been done \cite{chen2, zhang2, ergin2, valentine}. Generally, there are two types of design for these devices: 
isotropic and anisotropic carpet cloakings. While it is well-known that isotropic feature of a material is closely related to 
a conformal transformation, anisotropic property is related to a general coordinate transformation. In \cite{roman}, 
Schmied {\it et al.} proposed carpet cloaking devices and grating cloaks by using a Fourier transformation-like approach whose integral region 
is limited to positive value. 
 
In this work, inspired by \cite{roman}, we present a new method to design isotropic carpet cloaking using Laplace transformation. In particular, it 
enables us to systematically design a broad variety of carpet cloakings. Roughly speaking, each Laplace transformation 
will correspond to a new isotropic carpet cloaking.
 
The paper is organized as follows. First, we will introduce the methodology in Methods section. We then provide 
three examples of carpet cloaking designs computed using the Laplace transformation. Finally, we discuss the implications 
of our proposed methodology and conclude with the future work.

\section{Laplace transformation method for carpet cloaking}
Here, we will explain how to design a carpet cloaking by using Laplace transformation. Our construction 
is based on transformation optics \cite{report}. We consider a two dimensional cloaking device. The reason why we use 
two dimensions is that a 2-dimensional space can be identified with the complex plane and we can use complex analysis which is well-known.

Let $z\in \bf{C}$ be a complex number. A complex plane of $z$ is identified with a physical space. We call it $z$-space.
Let $w\in \bf{C}$ be also a complex number. A complex plane of $w$ will define a mathematical space. We call it $w$-space.
We then consider a Laplace transformation as follows:
\begin{eqnarray}\label{eqn:defining equation}
z=w+L[f(t)](w),
\end{eqnarray}
where $L[f(t)](w)$ is a Laplace transformation of function $f(t)$ defined by
\begin{eqnarray}
L[f(t)](w)=\int_0^{\infty}f(t)e^{-wt}dt.
\end{eqnarray}
Let us set $w=x+iy$, where $i^2=-1$. Though it is formal argument, when we take limit ($x\to \infty$), we obtain
\begin{eqnarray}
z\sim w,
\end{eqnarray}
where we assume that we can change limit operation and integral. This fact implies that in the large value region of real axis, 
the physical space and mathematical space are asymptotically identical and the light rays of mathematical 
space can be identically mapped into the physical space.   

In particular, the transformation law of refractive index between physical space and mathematical space is given by
\begin{eqnarray}
n_z=n_w|\frac{dw}{dz}|,
\end{eqnarray}
where $n_z$ is the refractive index of physical space and $n_w$ is the refractive index of mathematical space \cite{leon1,leon2,report}. Using equation (\ref{eqn:defining equation}), we obtain
\begin{eqnarray}\label{eqn:formula of refractive index}
n_z=\frac{n_w}{|1-L[tf(t)](w)|}.
\end{eqnarray}

\section{Application of the Laplace Transformation method}
Here, we will illustrate how to design a carpet cloaking using some examples of Laplace transformation.

\subsection{Case 1 ($f(t)=1$)}
\begin{figure}[htbp] 
\begin{minipage}{1\hsize}
  \begin{center}
  \includegraphics[scale=0.7]{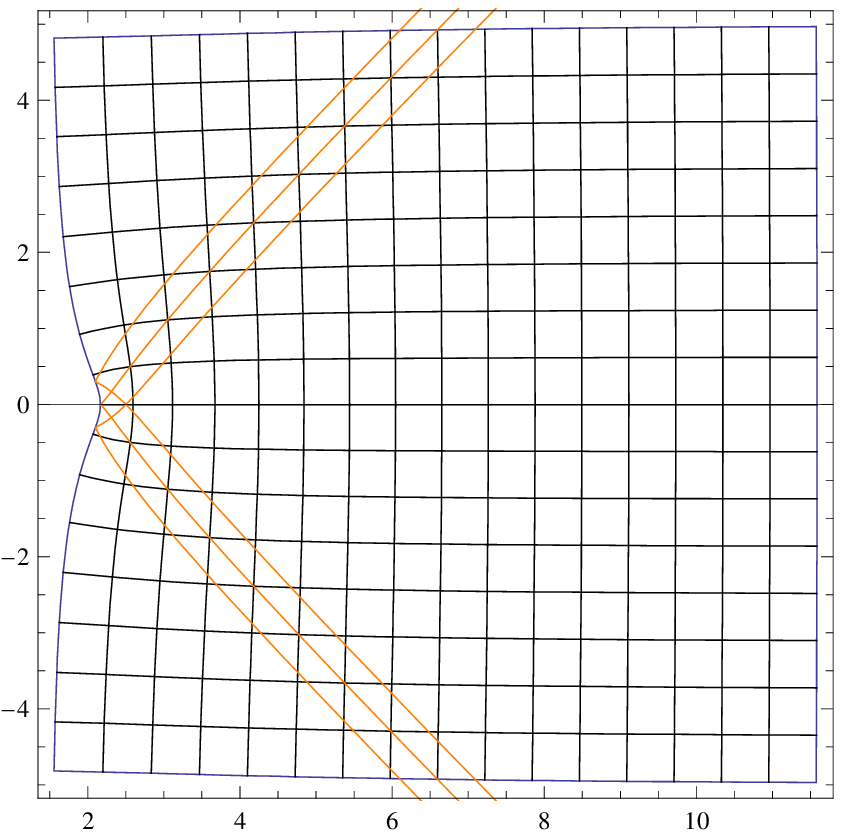} 
  \put(-180,170){(a)}
  \end{center}
 \end{minipage}
 \begin{minipage}{1\hsize}
  \begin{center}
   \includegraphics[scale=0.8]{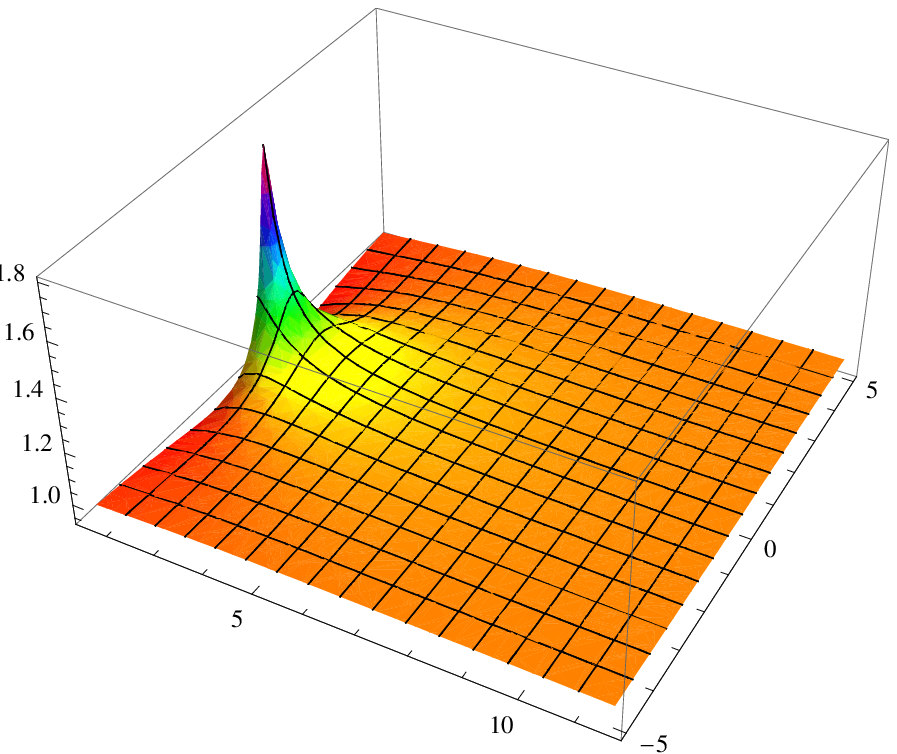}
   \put(-180,150){(b)}
  \end{center}  
 \end{minipage}
 \caption{$f(t)=1$ case. (a) Light rays in physical space $z$. The light rays are reflected at the bump. (b) Height axis 
 denotes refractive index of the cloaking device. In both figures, the boundary is set by using parameter $a=1.5$.}
 \label{fig:case 1}
\end{figure}

As an example, let us use the Laplace transformation of $f(t)=1$. Then, using Eq. (\ref{eqn:defining equation}), we obtain the following conformal map: 
\begin{eqnarray}\label{eqn: complex function of ex 1}
z=w+1/w.
\end{eqnarray}
We set the refractive index of $w$-space as $n_w$=1. Therefore all trajectory of light goes along straight lines. 
Using Eq. (\ref{eqn:formula of refractive index}), we obtain the refractive index as follows:
\begin{eqnarray} \label{eqn: refractive index of ex 1}
n_z=\frac{w^2}{w^2-1}.
\end{eqnarray}
The singular points of Eq. (\ref{eqn: complex function of ex 1}) and Eq. (\ref{eqn: refractive index of ex 1}) are 0 and $\pm1$ respectively. Therefore, both (\ref{eqn: complex function of ex 1}) and (\ref{eqn: refractive index of ex 1}) are convergent for $\mbox{Re}(w)>1$ and we can set the boundary of carpet cloakin
 in $w-$space as follows:
\begin{eqnarray}\label{eqn: boundary}
\mbox{Re}(w)=a ~~~~(a>1).
\end{eqnarray}
The physical shape of the bump and trajectory of light in $z$-space are obtained by pulling back the boundary (\ref{eqn: boundary}) 
and trajectory of light rays in $w$-space. We illustrate the trajectory of light rays and profile of refractive index 
of this example in Fig. \ref{fig:case 1}.

\subsection{Case 2 ($f(t)=\cos(2t)$)}
As the next example, we will use the Laplace transformation of $f(t)=\cos(2t)$.
Using equation (\ref{eqn:defining equation}), we obtain
\begin{eqnarray}\label{eqn: complex function of ex 2}
z=w+\frac{w}{w^2+4}.
\end{eqnarray}

\begin{figure}[htbp] 
\begin{minipage}{1\hsize}
  \begin{center}
  \includegraphics[scale=0.7]{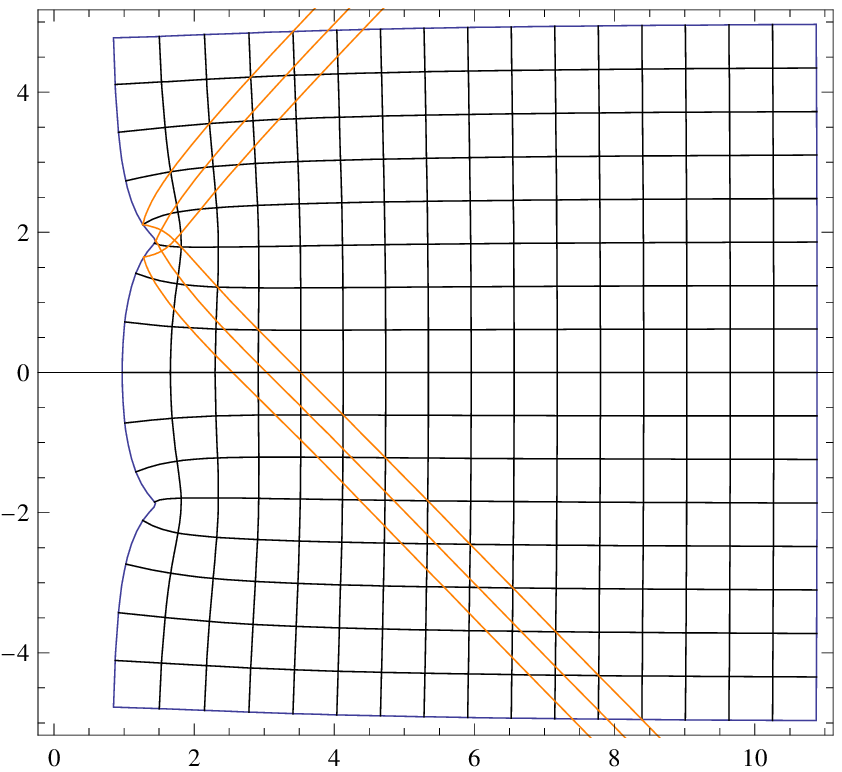} 
  \put(-180,170){(a)}
  \end{center}
 \end{minipage}
 \begin{minipage}{1\hsize}
  \begin{center}
   \includegraphics[scale=0.8]{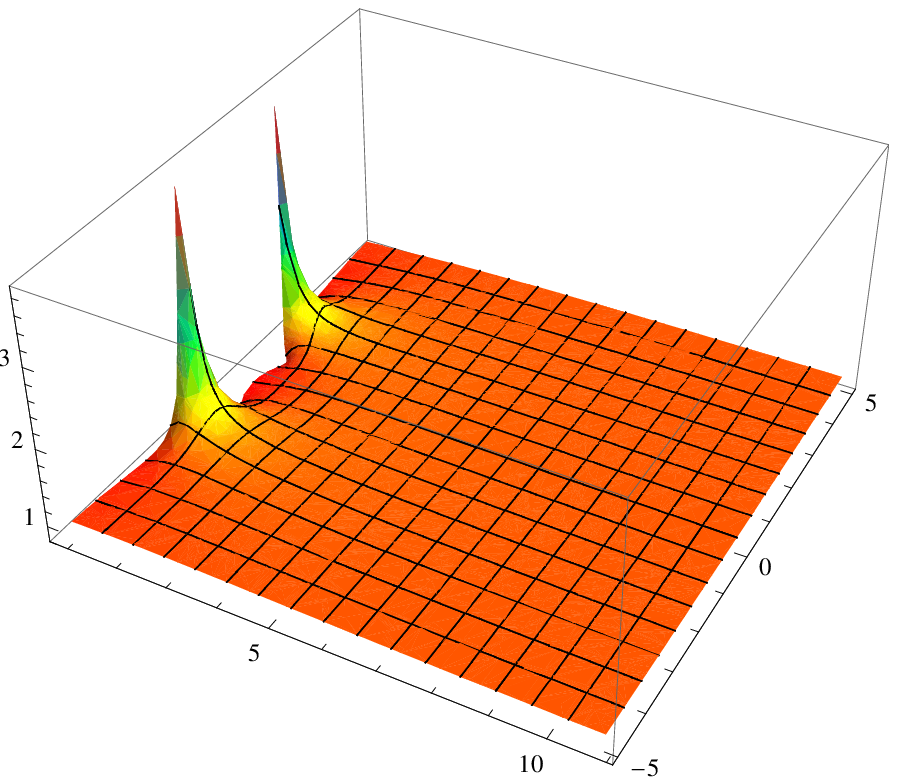}
   \put(-180,150){(b)}
  \end{center}  
 \end{minipage}
 \caption{$f(t)=\cos(2t)$ case. (a) Light rays in physical space $z$. The light rays are reflected at the bump. (b) Height axis 
 denotes refractive index of the cloaking device. In both figures, the boundary is set by using parameter $a=0.8.$}
\label{fig:case 2}
\end{figure}

The singular points of this equation are $w=\pm 2i$.
Applying Eq. (\ref{eqn:formula of refractive index}), we can compute the refractive index as follows:
\begin{eqnarray}\label{eqn: refractive index of ex 2}
n_z=\frac{(w^2+4)^2}{w^4+7w^2+20},
\end{eqnarray}
where we set $n_w=1$.
Refractive index (\ref{eqn: refractive index of ex 2}) has four singular points and they are located as follows:
\begin{eqnarray}
w\simeq \pm 0.697186 \pm i\times 1.99651.   
\end{eqnarray}
Both (\ref{eqn: complex function of ex 2}) and (\ref{eqn: refractive index of ex 2}) are convergent for $\mbox{Re}(w)>0.697186$ 
and we can set the boundary of carpet cloaking in $w$-space as follows:
\begin{eqnarray}
\mbox{Re}(w)=a ~~~~(a>0.697186).
\end{eqnarray}
We illustrate the trajectory of light rays and profile of refractive index of this example in Fig. \ref{fig:case 2}.

\subsection{Case 3 ($f(t)=\sin(2t)$)}
\begin{figure}[htbp] 
\begin{minipage}{1\hsize}
  \begin{center}
  \includegraphics[scale=0.7]{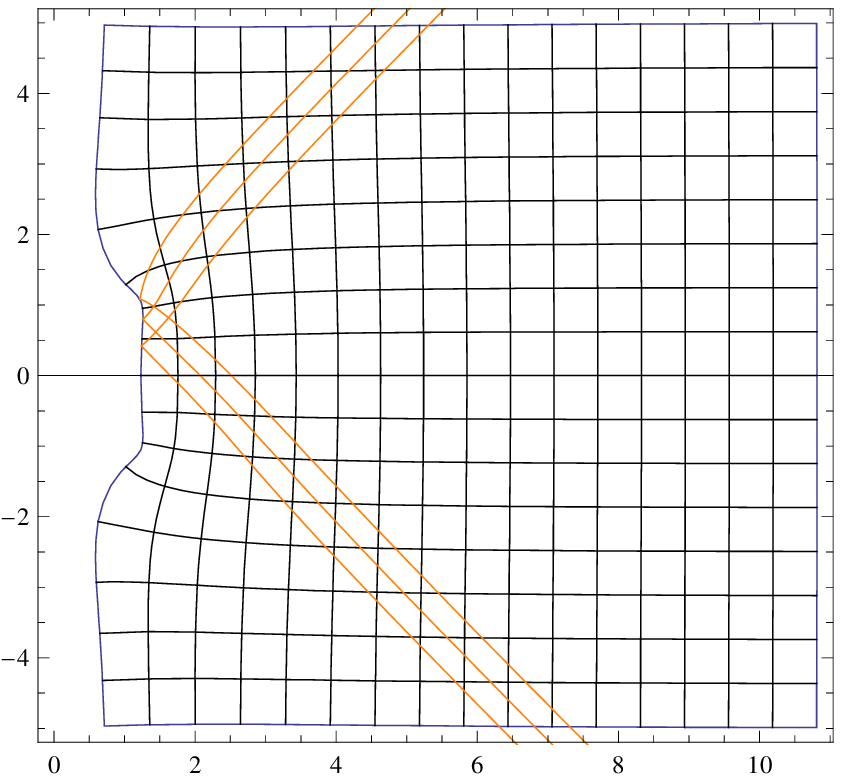} 
  \put(-180,170){(a)}
  \end{center}
 \end{minipage}
 \begin{minipage}{1\hsize}
  \begin{center}
   \includegraphics[scale=0.8]{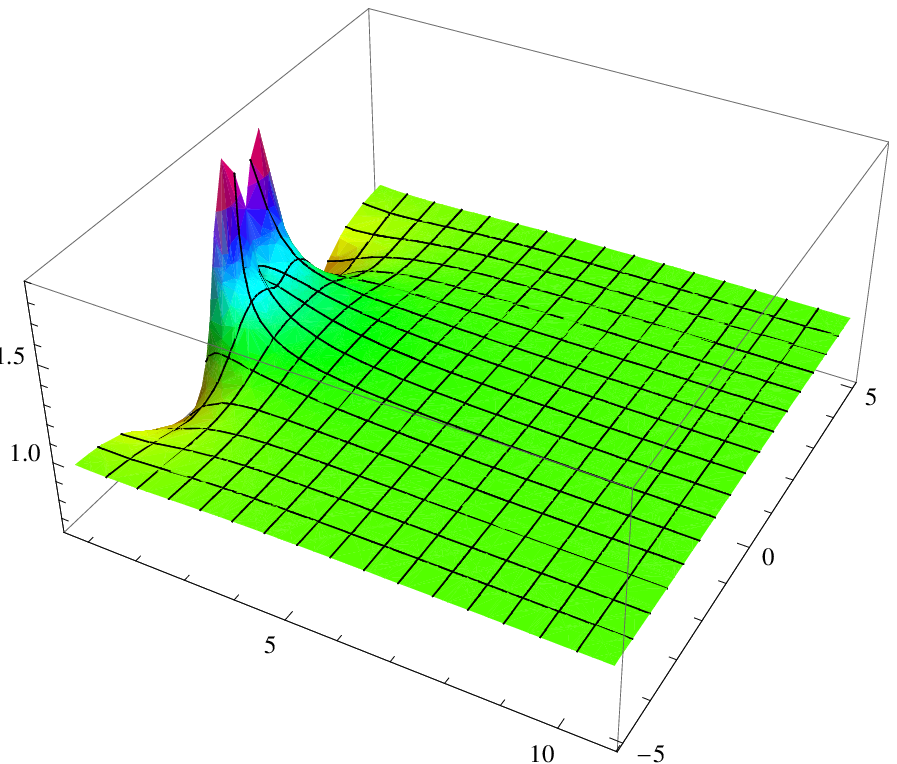}
   \put(-180,150){(b)}
  \end{center}  
 \end{minipage}
 \caption{Example ($f(t)=\sin(2t)$ case). (a) Light rays in physical space $z$. The light rays are reflected at the bump. (b) Height axis 
 denotes refractive index of the cloaking device. In both figures, the boundary is set by using parameter $a=0.8$.}
 \label{fig:case 3}
\end{figure}

As the third example, we use Laplace transformation of $f(t)=\sin(2t)$.
Using Eq. (\ref{eqn:defining equation}), we obtain 
\begin{eqnarray}\label{eqn: complex function of ex 3}
z=w+\frac{2}{w^2+4}.
\end{eqnarray}
The singular points are given by $w=\pm2i$.
Applying the transformation law (\ref{eqn:formula of refractive index}), we obtain the refractive index as follows:
\begin{eqnarray}\label{eqn: refractive index of ex 3}
n_z=\frac{(w^2+4)^2}{w^4+8w^2-4w+16},
\end{eqnarray}
where we set $n_w=1$.
Refractive index (\ref{eqn: refractive index of ex 3}) has four singular points and they are located as follows:
\begin{eqnarray}
w\simeq -0.49269 \pm i2.50448,\nonumber 
\end{eqnarray}
\begin{eqnarray}
w\simeq 0.49269 \pm i1.48764. 
\end{eqnarray}
Therefore, both (\ref{eqn: complex function of ex 3}) and (\ref{eqn: refractive index of ex 3}) are convergent for $\mbox{Re}(w) > 0.49269$. Thus, 
we can set the boundary of carpet cloaking in $w$-space as follows
\begin{eqnarray}
\mbox{Re}(w)=a ~~~~(a > 0.49269)
\end{eqnarray}
We illustrate the trajectory of light rays and profile of refractive index of this example in Fig. \ref{fig:case 3}.

\section{Discussion and Conclusion}

We have presented a new method to design carpet cloaking devices. This method allows us 
to systematically construct carpet cloaking devices by using a rich variety of functional forms $f(t)$. Therefore, we can potentially hide
a large variety of bumps with different geometrical shapes just by selecting a suitable $f(t)$ function. The method is simple and allows us
to derive the required refractive index profile for each specific carpet cloaking in a straightforward way, without having to consider 
deeply the complex mathematics involved in the transformation optics related to cloaking devices. 

In \cite{roman}, Schmied {\it et al.} constructed a carpet cloaking device using Fourier transformation-like approach, 
whose integral region is limited to only positive values. As a consequence, it is not straightfoward to compute the integral, because the 
integration region of their transformation is not the entire real axis and they could not use well-known standard Fourier transformation techniques. 
In contrast, our proposed method provides a direct connection between carpet cloaking and Laplace transformation, so that we can 
use established mathematical knowledge of Laplace transformation for designing a carpet cloaking. As a result, 
it makes easier the design of a carpet cloaking.

Moreover, in sharp constrast to the cloaking device problems where refractive index profiles are required 
to take extreme positive or even negative values, the examples of carpet cloaking shown in this work indicate that the required refractive 
index profiles are within reasonable values and that can be found
in already existing materials, or be possibly constructed with metamaterial state-of-art technologies. This finding, therefore, shows that
the carpet cloaking devices are not only a theoretical design but also could be realized with current materials science and technology.  

As a future work, we may investigate the physical meaning of the functional form $f(t)$ and its relation with each specific 
geometrical shape of the bump. It could be useful to have a mathematical relationship that links that geometry of the bump with this function. 
It would then allow us to design carpet cloakings to conceal more complex geometric structures. On the other hand, another theoretical 
issue that requires attention would be the extension of the current methods to 3-dimensional problems. In the meantime, the experimental work on 
carpet cloaking in 2-dimensional spaces should be encouraged.

\begin{acknowledgments}
T.O. and J.C.N. gratefully acknowledge the funding support of a Grant-in-Aid for Scientific
Research (C) from MEXT, Japan.
\end{acknowledgments}

\subsection{}
\subsubsection{}

\bibliography{basename of .bib file}

\end{document}